# Bloch sphere model for two-qubit pure states

Chu-Ryang Wie[1]

State University of New York at Buffalo, Department of Electrical Engineering, 230B Davis Hall, Buffalo, NY 14260

The two-qubit pure state is explicitly parameterized by three unit 2-spheres and a phase factor. For separable states, two of the three unit spheres are the Bloch spheres of each qubit with coordinates ($\theta_A$,$\phi_A$) and ($\theta_B$,$\phi_B$). The third sphere parameterizes the degree and phase of concurrence, an entanglement measure. This sphere may be considered a 'variable' complex imaginary unit ***t*** where the stereographic projection maps the qubit-A Bloch sphere to a complex plane with this variable imaginary unit. This Bloch sphere model gives a consistent description of the two-qubit pure states for both separable and entangled states. We argue that the third sphere (entanglement sphere) parameterizes the nonlocal properties, entanglement and a nonlocal relative phase, while the local relative phases are parameterized by the azimuth angles, $\phi_A$ and $\phi_B$, of the two quasi-Bloch spheres. On these three unit spheres and a phase factor ($\zeta_B$), the two-qubit pure states and unitary gates can be geometrically represented. Accomplished by means of Hopf fibration, the complex amplitudes ($\alpha$, $\beta$, $\gamma$, $\delta$) of a two-qubit pure state and the Bloch sphere parameters are related by a single quaternionic relation:

$$\begin{pmatrix} \alpha + \beta j \\ \gamma + \delta j \end{pmatrix} = \begin{pmatrix} cos\dfrac{\theta_A}{2} \\ sin\dfrac{\theta_A}{2} e^{t\phi_A} \end{pmatrix} \left( \cos\frac{\theta_B}{2} + \sin\frac{\theta_B}{2} e^{k\phi_B} j \right) e^{k\zeta_B}$$

## I. INTRODUCTION

The single qubit Bloch sphere provides a useful means of visualizing the state and a neat description of many single qubit operations, enabling physical intuitions and serving as an excellent testbed for ideas about quantum computation and information [1]. However there is no simple generalization of Bloch sphere for multiple qubits, not even for two qubits. Much that is weird and wonderful about quantum mechanics can be appreciated by considering the properties of the quantum states of two qubits [2]. Any multi-qubit quantum logic gates can be built out of a two-qubit operation and a number of single-bit operations [1-3]. All of the above points give a strong reason to develop a practical Bloch sphere-like tool for the two-qubit states and gates. In spite of a considerable amount of efforts spent to develop a feasible two-qubit Bloch sphere model [4-7], there still is no Bloch sphere model that is simple enough to serve as useful a tool as the single qubit Bloch sphere does for state representation and manipulation.

For complete description, two qubit states require seven parameters for pure states and fifteen parameters for general mixed states. The two qubit state space is usually described by a unit 7-sphere $S^7$ for pure states and by a four-dimensional special unitary group SU(4) for a mixed state density matrix and two-qubit operators. There have been various attempts to parameterize the two qubit state space, including an explicit parameterization of SU(4) [4], and construction of special unitary group on two qubit Hilbert space using geometric algebra of a six-dimensional real Euclidean vector space [5]. Havel and Doran presented, as a special case to their discussion of SU(4) mixed state model, an explicit parameterization of a two-qubit pure state with seven angle parameters which was written as a convex combination of tensor products of one-qubit states [5]. Mosseri and Dandoloff applied Hopf fibration and attempted to develop a Bloch sphere model for two-qubit pure state [6,7]. They proposed a 3-dimensional ball with equal-concurrence concentric spherical shells as a possible model, which obviously has far too few degrees of freedom to specify the two-qubit states. However, their approach with Hof fibration has a close connection to our work in this paper.

In this paper, we limit to the two-qubit pure states and attempt to find a Bloch sphere model that is simple and practical. We parameterize the $S^7$ two-qubit pure state space by 7 angle parameters which are grouped into the two Bloch coordinates ($\theta_A$, $\phi_A$) and ($\theta_B$, $\phi_B$), the entanglement and its relative phase ($\chi$, $\xi$), and a phase angle ($\zeta_B$). These seven angle parameters are related to the two-qubit quantum state amplitudes, α, β, γ and δ via a single equation. The rest of this article is organized as follows: In section II, we briefly review the Hopf map with the non-commutative quaternion algebra. The Hopf fibration maps the $S^7$ two-qubit pure state space to an $S^4$ base space. This is done by mapping to each point of the $S^4$ base space the entire $S^3$ fiber space. This Hopf map enables us to model a quantum state in the $S^7$ space by a product of a function in the $S^4$ base space and a function in the $S^3$ fiber space. In section III, we define the three unit 2-spheres and a phase factor to serve as the Bloch sphere model for two-qubit pure states. In section IV, we present a streamlined procedure for, given the four complex state amplitudes (α, β, γ and δ), how to find the seven angle parameters of the Bloch spheres, and vice versa. In this section we also show some concrete examples of Bloch coordinates of the maximally entangled Bell states and several separable states, and examples of the two-qubit quantum gates, CNOT, CZ and SWAP gates. In section V, we discuss the entanglement sphere parameters (χ, ξ), phase angle parameter ($\zeta_B$), and the mixed-state single-qubit Bloch ball after partial trace and its relation to the quasi-Bloch sphere. In section VI, we give a summary conclusion.

[1] Electronic address: wie@buffalo.edu

## II. THE HOPF FIBRATION OF TWO-QUBIT STATES AND PARAMETERIZATION

For two-qubit pure states,

$$\left|\Psi_{AB}\right\rangle = \alpha\left|00\right\rangle + \beta\left|01\right\rangle + \gamma\left|10\right\rangle + \delta\left|11\right\rangle \qquad (1)$$

where,

$$\alpha,\beta,\gamma,\delta \in \mathbb{C} \text{ and } \left|\alpha\right|^2 + \left|\beta\right|^2 + \left|\gamma\right|^2 + \left|\delta\right|^2 = 1,$$

our Bloch sphere representation starts by introducing quaternion 'amplitudes' by defining them as follows:

$$q_{0A} \equiv \alpha + \beta j,\;\; q_{1A} \equiv \gamma + \delta j \qquad (2)$$

or

$$q_{0B} \equiv \alpha + \gamma j,\;\; q_{1B} \equiv \beta + \delta j \qquad (3)$$

where, the quaternion imaginary units *i, j* and *k* satisfy the following multiplication rules and identities:

$$\begin{aligned} &i^2 = j^2 = k^2 = ijk = -1,\;\; i = jk = -kj, \\ &j = ki = -ik,\;\; k = ij = -ji. \end{aligned} \qquad (4)$$

Throughout this paper, *k* will be assumed to be the imaginary unit of ordinary complex numbers. A quaternion forms non-commutative algebra and may be written in terms of two complex numbers as in Eqs.(2) and (3), or in terms of four real numbers, *a, b, c* and *d* as

$$q = a + bi + cj + dk,\; a,b,c,d \in \mathbb{R} \qquad (5)$$

A pure imaginary unit quaternion squares to -1 like the three imaginary units in Eq.(4) and thus, a pure unit quaternion *t* provides the same relation between an exponential function and trigonometric (sine and cosine) functions as the complex imaginary unit does, and it can be parameterized in terms of two real numbers. That is,

$$\begin{aligned} &t^2 = -1;\; e^{t\phi} = cos\phi + tsin\phi; \\ &t = i\,sin\chi cos\xi + j\,sin\chi sin\xi + k\,cos\chi;\;\; \phi,\chi,\xi \in \mathbb{R} \end{aligned}. \qquad (6)$$

The normalization condition

$$\left|q_{0A}\right|^2 + \left|q_{1A}\right|^2 = \left|q_{0B}\right|^2 + \left|q_{1B}\right|^2 = \left|\alpha\right|^2 + \left|\beta\right|^2 + \left|\gamma\right|^2 + \left|\delta\right|^2 = 1 \qquad (7)$$

identifies the state space of a two-qubit pure state as a unit 7-sphere, $S^7$, which is embedded in $\mathbb{R}^8$.

The Hopf map from the two-qubit state space $S^7$ to the base $S^4$ is a composition of two maps: first from $S^7$ to $\mathbb{R}^4$ (+∞), then to $S^4$ by an inverse stereographic projection [7, 8]. An explicit treatment of conversion from the complex amplitudes, α, β, γ and δ, to the quaternion 'amplitudes' $q_0$ and $q_1$ is useful in relating the target space of the Hopf map to each qubit space (and thus identifying the Bloch parameters of each space). We do this as follows:

### A. In the basis of qubit-A

$$\left|\Psi_{AB}\right\rangle = \left|0\right\rangle_A \otimes \left(\alpha\left|0\right\rangle_B + \beta\left|1\right\rangle_B\right) + \left|1\right\rangle_A \otimes \left(\gamma\left|0\right\rangle_B + \delta\left|1\right\rangle_B\right)$$

$$\longrightarrow$$

$$\left|\tilde{\Psi}_A\right\rangle = \left|0\right\rangle_A\left(\alpha + \beta j\right) + \left|1\right\rangle_A\left(\gamma + \delta j\right) = \begin{pmatrix} \alpha + \beta j \\ \gamma + \delta j \end{pmatrix} = \begin{pmatrix} q_{0A} \\ q_{1A} \end{pmatrix} \qquad (8)$$

where, after the Hopf fibration, qubit-A is identified with the base space $S^4$ and the qubit-B with the fiber space $S^3$.

The above conversion process keeps the qubit-A basis vectors intact while identifying the qubit-B basis vectors with the quaternion units: $\left|0\right\rangle_B \leftrightarrow 1,\; \left|1\right\rangle_B \leftrightarrow j$. This correspondence is applied again when recovering the qubit-B state at the end of calculation. Let us define a density matrix-like quantity as follows:

$$\tilde{\rho}_A \equiv \left|\tilde{\Psi}_A\right\rangle\left\langle\tilde{\Psi}_A\right| = \begin{pmatrix} q_{0A} \\ q_{1A} \end{pmatrix}\left(\bar{q}_{0A}, \bar{q}_{1A}\right) = \begin{pmatrix} \alpha+\beta j \\ \gamma+\delta j \end{pmatrix}\left(\bar{\alpha}-\beta j, \bar{\gamma}-\delta j\right) \tag{9}$$

where the over-bar indicates the quaternion or complex conjugation which reverses the sign of imaginary units: $\bar{i} = -i$, $\bar{j} = -j$, and $\bar{k} = -k$. As will be shown below, all local parameters of the qubit-B are cancelled out in this dyad $\left|\tilde{\Psi}_A\right\rangle\left\langle\tilde{\Psi}_A\right|$; and the only parameters remaining in $\tilde{\rho}_A$ are the Bloch coordinates of the base qubit (qubit-A) and two nonlocal parameters (entanglement and phase). These four parameters (of the Hopf base) are determined from the products of the complex amplitudes α, β, γ and δ. From Eq.(9) we obtain

$$\tilde{\rho}_A = \rho_A + \left(\alpha\delta-\beta\gamma\right)\begin{pmatrix} 0 & -j \\ j & 0 \end{pmatrix}, \text{ where}$$

$$\rho_A = \begin{pmatrix} |\alpha|^2+|\beta|^2 & \alpha\bar{\gamma}+\beta\bar{\delta} \\ \bar{\alpha}\gamma+\bar{\beta}\delta & |\gamma|^2+|\delta|^2 \end{pmatrix} \tag{10}$$

Here, $2\left|\alpha\delta-\beta\gamma\right|$ is the concurrence, an entanglement measure [9], and $\rho_A$ is the reduced density matrix for qubit-A given by $\rho_A = tr_B\left(\left|\Psi_{AB}\right\rangle\left\langle\Psi_{AB}\right|\right)$. It is worth noting that the off-diagonal element of $\rho_A$ represents the coherence of qubit-A after qubit-B is traced out, and Eq.(10) shows that the other off-diagonal element of $\tilde{\rho}_A$ represents the entanglement of the composite state.

### B. In the basis of qubit-B

We write the bipartite state in the qubit-B basis by identifying the qubit-A basis vectors with the quaternion units *1* and *j* as follows:

$$\left|\Psi_{AB}\right\rangle = \left(\alpha|0\rangle_A + \gamma|1\rangle_A\right)\otimes|0\rangle_B + \left(\beta|0\rangle_A + \delta|1\rangle_A\right)\otimes|1\rangle_B$$

$$\longrightarrow$$

$$\left|\tilde{\Psi}_B\right\rangle = \left(\alpha+\gamma j\right)|0\rangle_B + \left(\beta+\delta j\right)|1\rangle_B = \begin{pmatrix} \alpha+\gamma j \\ \beta+\delta j \end{pmatrix} = \begin{pmatrix} q_{0B} \\ q_{1B} \end{pmatrix} \tag{11}$$

where, after the Hopf fibration, qubit-B will be in the base space $S^4$ and qubit-A in the fiber space $S^3$.

The quasi density matrix is

$$\begin{aligned} \tilde{\rho}_B &\equiv \left|\tilde{\Psi}_B\right\rangle\left\langle\tilde{\Psi}_B\right| = \begin{pmatrix} \alpha+\gamma j \\ \beta+\delta j \end{pmatrix}\left(\bar{\alpha}-\gamma j, \bar{\beta}-\delta j\right) \\ &= \rho_B + \left(\alpha\delta-\beta\gamma\right)\begin{pmatrix} 0 & -j \\ j & 0 \end{pmatrix}, \\ \rho_B &= \begin{pmatrix} |\alpha|^2+|\gamma|^2 & \alpha\bar{\beta}+\gamma\bar{\delta} \\ \bar{\alpha}\beta+\bar{\gamma}\delta & |\beta|^2+|\delta|^2 \end{pmatrix} \end{aligned} \tag{12}$$

Where $\rho_B$ is the reduced density matrix of qubit-B: $\rho_B = tr_A\left(\left|\Psi_{AB}\right\rangle\left\langle\Psi_{AB}\right|\right)$. It is interesting to note that whether we choose qubit-A basis or qubit-B basis, the resulting quasi density matrix $\tilde{\rho}$ consists of the reduced density matrix of the base qubit plus an identical term representing the concurrence. Also the quasi-density matrix $\tilde{\rho}$ possesses the pure-state density matrix properties: $\tilde{\rho}^2 = \tilde{\rho}$, tr($\tilde{\rho}$)=1, and det($\tilde{\rho}$)=0. A two-qubit SWAP operation will send the quasi state $\left|\tilde{\Psi}_A\right\rangle$ to $\left|\tilde{\Psi}_B\right\rangle$, or the quasi-density matrix $\tilde{\rho}_A$ to $\tilde{\rho}_B$, and vice versa. The trajectory of a SWAP gate on the Bloch spheres will be shown later.

### C. Hopf fibration

Before we parameterize the bipartite pure state space $S^7$, we briefly review the Hopf fibration. The Hopf fibration consists of a composition of two maps: the Hopf map $h_1$ which maps $S^7$ to $\mathbb{R}^4$, followed by an inverse stereographic projection $h_2$ which maps $\mathbb{R}^4$ to $S^4\backslash(1,0,0,0,0)$.

$$h_1: \underbrace{\begin{pmatrix} q_0 \\ q_1 \end{pmatrix}}_{S^7} \rightarrow \underbrace{\overline{q_0 q_1^{-1}}}_{\mathbb{R}^4} \equiv Q = Q_1 + Q_2 i + Q_3 j + Q_4 k \tag{13}$$

$$q_0, q_1, Q \in \mathbb{H}; \; |q_0|^2 + |q_1|^2 = 1; \; Q_i \in \mathbb{R}; \quad i = 1,2,3,4$$

$$h_2: \underbrace{\left(Q_1, Q_2, Q_3, Q_4\right)}_{\mathbb{R}^4} \rightarrow \underbrace{\left(x_0, x_1, x_2, x_3, x_4\right)}_{S^4} \tag{14}$$

$$\sum_{i=0}^{4} x_i^2 = 1, \quad x_i \in \mathbb{R},$$

The stereographic projection gives

$$\overline{q_0 q_1^{-1}} = Q = \left(Q_1, Q_2, Q_3, Q_4\right) = \left(\frac{x_1}{1-x_0}, \frac{x_2}{1-x_0}, \frac{x_3}{1-x_0}, \frac{x_4}{1-x_0}\right) \tag{15}$$

which maps $S^4\backslash\left(1,0,0,0,0\right)$ to $\mathbb{R}^4$, or vice versa. This is illustrated in Fig.1.

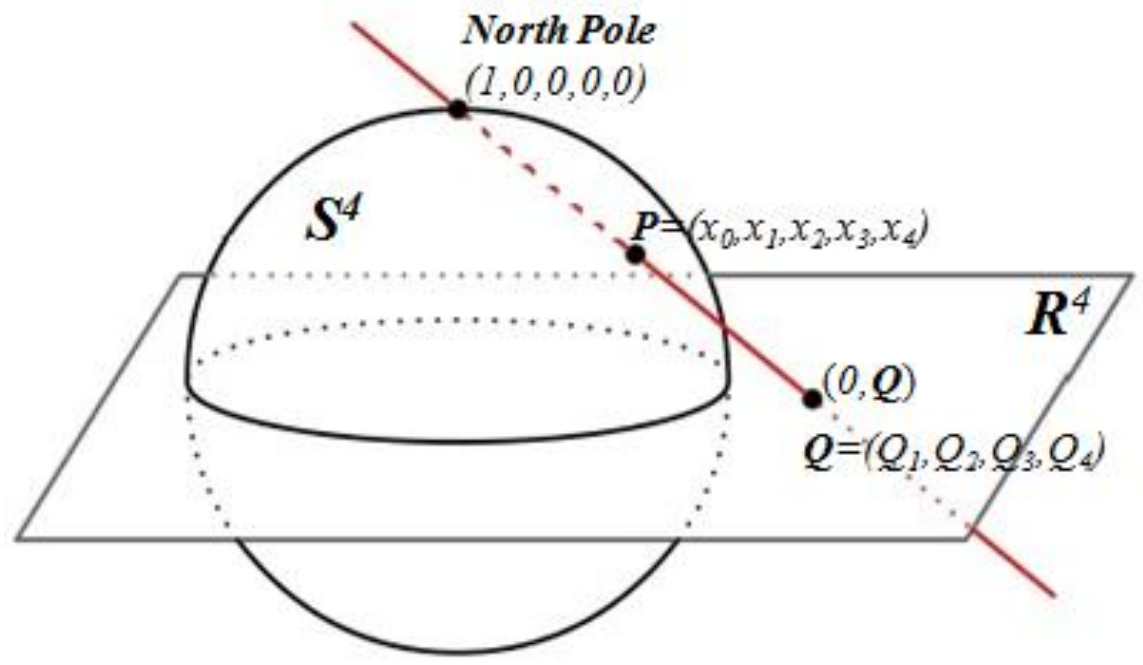

Figure 1 Stereographic projection from $S^4$ to $\mathbb{R}^4$ where the coordinates of $S^4$ are related to the coordinates of $\mathbb{R}^4$ through Eq.(15).

Therefore, a pair of quaternions, $q_0$ and $q_1$, are mapped to the coordinate ($x_0$, $x_1$, $x_2$, $x_3$, $x_4$) on the 4-sphere $S^4$ which is embedded in $\mathbb{R}^5$ with the North pole excluded. This Hopf fibration reads $h: S^7 \xrightarrow{S^3} S^4$ where $h = h_2 \circ h_1$. An inspection shows that all the pairs of quaternions ($q_0$, $q_1$), which differ only by a right multiplication of an arbitrary unit quaternion $q_f$, will map to the same point Q in $\mathbb{R}^4$.

$$Q = \frac{q_1\bar{q}_0}{|q_1|^2} = \frac{(q_1 q_f)\left(\overline{q_0 q_f}\right)}{|q_1 q_f|^2} \quad \text{where}$$

$$|q_f|^2 = q_f\bar{q}_f = \bar{q}_f q_f = 1 \qquad (16)$$

Therefore,

$$h_1 : \begin{pmatrix} q_0 q_f \\ q_1 q_f \end{pmatrix} \to Q \quad \text{for any unit quaternion } q_f \qquad (17)$$

This unit quaternion $q_f$ is the $S^3$ fiber in the Hopf map $h: S^7 \xrightarrow{S^3} S^4$. It is worth repeating here that the Hopf map *h* maps an entire subspace $S^3$ of $S^7$ to a single point of the base space, the 4-sphere $S^4$. The unit quaternion $q_f$ is itself fibrated by another Hopf map $h_3$: $S^3 \xrightarrow{S^1} S^2$ [8], which maps an entire circle $S^1$, a subspace of $S^3$, to a single point on $S^2$.

$$h_3 : \overset{S^3}{q_f} \to \overset{S^2}{q_f k \bar{q}_f} \qquad (18)$$

where *k* is an imaginary unit. The image of the map $h_3$, $q_f k \bar{q}_f$, is another pure-imaginary unit quaternion which is obtained by rotating *k* around an axis of rotation given by the imaginary part of $q_f$ by an angle of rotation, $2cos^{-1}(Re(q_f))$, given by the real part of $q_f$ [8]. A pure unit quaternion can be parameterized by a unit 2-sphere $S^2$ and will form the base (or the image space) of the Hopf map $h_3$. Its fiber $S^1$ is a unit circle which can be represented by a unit complex number: $e^{k\zeta}$.

### D. Parameterization of the $S^4$ Hopf base and the $S^3$ Hopf fiber

The above discussion of Hopf fibration of the $S^7$ two-qubit pure state space suggests a parameterization in terms of two groups of parameters: the $S^4$ base space parameters and the $S^3$ fiber space parameters. We first parameterize the $S^4$ Hopf base as follows:

$$\overset{S^7}{\begin{pmatrix} q_0 \\ q_1 \end{pmatrix}} = \overbrace{\begin{pmatrix} cos\dfrac{\theta}{2} \\ sin\dfrac{\theta}{2} e^{t\phi} \end{pmatrix}}^{S^4} \overset{S^3}{q_f} \qquad (19)$$

$$\begin{aligned} t &= i\, sin\chi cos\xi + j\, sin\chi sin\xi + k\, cos\chi \\ &\equiv t_x i + t_y j + t_z k; \;\; \theta, \phi, \chi, \xi \in \mathbb{R} \end{aligned} \;, \qquad (20)$$

where $q_f$ is the unit quaternion, representing the $S^3$ Hopf fiber. By substituting Eqs.(19) and (20) into Eq.(15), we find

$$\begin{aligned} & x_0 = cos\theta, x_1 = sin\theta cos\phi, \; x_2 = sin\theta sin\phi sin\chi cos\xi = \\ & c\, cos\xi = bt_x, x_3 = sin\theta sin\phi sin\chi sin\xi = c\, sin\xi = bt_y, \\ & x_4 = sin\theta sin\phi cos\chi = b\, cos\chi = bt_z, \\ & b \equiv sin\theta sin\phi, c \equiv sin\theta sin\phi sin\chi = b\, sin\chi \end{aligned} \qquad (21)$$

Here, the $S^4$ Hopf base is parameterized in terms of two angles θ and ϕ and a pure-imaginary unit quaternion ***t*** which is parameterized by another two angles χ and ξ. The Cartesian coordinate of $S^4$ embedded in $\mathbb{R}^5$ is ($x_0$, $x_1$, $x_2$, $x_3$, $x_4$), as shown in Fig.1 and given in Eq.(21) in terms of these four angle parameters. We may regard the pure-imaginary unit quaternion ***t*** as a 'variable' complex imaginary unit for the complex plane onto which a 2-sphere $S^2$ with (θ, ϕ) spherical coordinate is mapped by a stereographic projection. Hence, the Hopf base $S^4$ consists of a unit 2-sphere $S^2$ (parameters: θ, ϕ) sitting atop another $S^2$ (parameters: χ, ξ). The latter $S^2$ represents the 'variable' imaginary unit ***t***. Also, a pure-imaginary quaternion can be viewed as a vector in a 3-dimensional space with the orthogonal axes defined by imaginary units *i, j, k*. The angles, $\theta$ and $\phi$, define the Bloch coordinates of the qubit in the base space. The pure unit quaternion ***t*** embodies the nonlocal properties such as concurrence and its phase. This will be discussed further in the next section. The interpretation of the $S^4$ base as being a 2-sphere $S^2$ sitting on top of another 2-sphere $S^2$ is also suggested by the following relation:

$$\underbrace{x_0^{\;2}+x_1^{\;2}+x_2^{\;2}+x_3^{\;2}+x_4^{\;2}}_{S^4}=\underbrace{x_0^{\;2}+x_1^{\;2}+b^2}_{S^2}\left(\underbrace{t_x^{\;2}+t_y^{\;2}+t_z^{\;2}}_{S^2}\right)=1 \qquad (22)$$

This equation suggests that the $S^4$ Hopf base can be parameterized in terms of two independent $S^2$ spheres. The first $S^2$ is in the $x_1$-$b$-$x_0$ coordinate frame (the base qubit Bloch sphere), and the second $S^2$ is in the $t_x$-$t_y$-$t_z$ (or $i$-$j$-$k$) coordinate frame (of the nonlocal parameter $\boldsymbol{t}$).

The unit quaternion $q_f$ representing the Hopf fiber enters Eq.(19) through a right multiplication to the base quaternions, and because the quaternion multiplication is non-commutative, this relative position is important. The unit quaternion $q_f$ is parameterized as

$$\overset{S^3}{q_f}=(\overbrace{\cos\frac{\theta_f}{2}+\sin\frac{\theta_f}{2}e^{k\phi_f}j}^{S^2})\overset{S^1}{e^{k\zeta_f}}\;;\;\;\theta_f,\phi_f,\zeta_f\in\mathbb{R} \qquad (23)$$

According to the Hopf map defined in Eq.(18), a given point ($\theta_f$, $\phi_f$) on the base space $S^2$ has a preimage which is a great circle on the 3-sphere $S^3$, and this great circle is parameterized by a unit complex number $exp(k\zeta_f)$. This great circle is also called the Hopf circle and is the 1-sphere $S^1$ of Eq.(23). In order to correctly parameterize the great circles of $S^3$, the Hopf fiber $S^1$, $exp(k\zeta_f)$, must appear at the right end of the expression in Eq(23), and once again, this relative position is important because quaternion multiplication is not commutative. The Hopf fiber $e^{k\zeta_f}$ not only provides the global phase of the bipartite state but it affects both the relative phase of qubit-B state and the azimuth angle $\xi$ of $t$. This will be shown explicitly below.

In summary, the angle parameters are: Hopf base $S^4$ ($\theta$, $\phi$, $\chi$, $\xi$) and Hopf fiber $S^3$ ($\theta_f$, $\phi_f$, $\zeta_f$) which is further mapped to $S^2$ base ($\theta_f$, $\phi_f$) and $S^1$ fiber ($\zeta_f$).

## III. TWO-QUBIT BLOCH SPHERE MODEL

In the remainder of this paper we discuss parameterization with respect to the qubit-A basis. The Hopf map, $h:S^7\xrightarrow{S^3}S^4$ and the parameterization Eq.(19), and the Hopf map $h_3$: $S^3\xrightarrow{S^1}S^2$ and the parameterization Eq.(23), are combined as

$$\left|\tilde{\Psi}_{\mathrm{A}}\right\rangle=\begin{pmatrix}\alpha+\beta j\\ \gamma+\delta j\end{pmatrix}=\begin{pmatrix}cos\dfrac{\theta_{\mathrm{A}}}{2}\\ sin\dfrac{\theta_{\mathrm{A}}}{2}e^{t\phi_A}\end{pmatrix}\left(\cos\frac{\theta_B}{2}+\sin\frac{\theta_B}{2}e^{k\phi_B}j\right)e^{k\zeta_B} \qquad (24)$$

The Hopf base ($S^4$) coordinates ($x_1$, $x_2$, $x_3$, $x_4$, $x_0$) in $R^5$ are separated into two $S^2$ Cartesian coordinates, ($x_1$, b, $x_0$) and ($t_x$, $t_y$, $t_z$), according to:

$x_1^{\;2}+x_2^{\;2}+x_3^{\;2}+x_4^{\;2}+x_0^{\;2}=x_1^{\;2}+b^2(t_x^{\;2}+t_y^{\;2}+t_z^{\;2})+x_0^{\;2}=1$

For the $S^4$ Hopf base, the first $S^2$ is qubit-A Bloch sphere with $x_1$-$b$-$x_0$ axis, and the second $S^2$ parameterizes the global parameter $\boldsymbol{t}$ with the degree of entanglement and its phase on the $t_x$-$t_y$-$t_z$ (or equivalently, $i$-$j$-$k$) axis. The qubit-B Bloch sphere is identical to a single qubit Bloch sphere coordinate as it arises from the Hopf fiber, a unit quaternion $q_B$ represented by 3-sphere $S^3$ which is further mapped by $h_3$ to the $S^2$ base (Bloch sphere, coordinates $\theta_B$ and $\phi_B$) and the $S^1$ fiber (a phase factor, exp($k\zeta_B$)). See Eqs.(23) and (24). Therefore, the $S^7$ pure two-qubit states are represented by three unit 2-spheres ($S^2$) and one unit circle $S^1$.

**<u>Qubit-A quasi-Bloch Sphere $S^2$</u> ($\theta_A$, $\phi_A$):**

$x_1=sin\theta_A cos\phi_A$, $b=sin\theta_A sin\phi_A$, $x_0=cos\theta_A$.

For general, entangled states, the qubit-A quasi-Bloch coordinate is ($\theta_A$, $\phi_A$). For separable states, the qubit-A Bloch coordinate is also ($\theta_A$, $\phi_A$). [$c=bsin\chi$; $x_2=c\ cos\xi$; $x_3=c\ sin\xi$; $x_4=b\ cos\chi$.]

**<u>Entanglement Sphere $S^2$</u> ($\chi$, $\xi$):** $t_x=sin\chi cos\xi$, $t_y=sin\chi sin\xi$, $t_z=cos\chi$; $\boldsymbol{t}=t_x i+t_y j+t_z k$.

For separable states (zero concurrence, $c=bsin\chi=0$), either the qubit-A Bloch coordinate is at the great circle in the $x_0x_1$-plane (i.e., $b=0$), or the entanglement sphere coordinate is at its North pole ($\chi=0$, $t=k$) or at South pole ($\chi=\pi$, $t=-k$). The phase factor $e^{k\zeta_B}$ influences $\xi$.

**<u>Qubit-B quasi-Bloch Sphere $S^2$</u> ($\theta_B$, $\phi_B$):**

$x_B=sin\theta_B cos\phi_B$, $y_B=sin\theta_B sin\phi_B$, $z_B=cos\theta_B$

For general states, the qubit-B quasi-Bloch coordinate is ($\theta_B$, $\phi_B$). The phase factor $e^{k\zeta_B}$ influences $\phi_B$. But ($\theta_B$, $\phi_B$) is used in plotting the Bloch sphere coordinate in this paper.

**<u>The Phase $S^1$</u>:** $e^{k\zeta_B}$

$S^1$($e^{k\zeta_B}$) is the Hopf fiber of $q_B$ and it must appear at the right-most end of the expression for quaternion $q_B$, as shown in Eq.(25) below. This phase factor may be considered the global phase only after properly adjusting (i.e., shifting by $2\zeta_B$) the azimuths $\xi$ and $\phi_B$.

$$q_B=\left(\cos\frac{\theta_B}{2}+\sin\frac{\theta_B}{2}e^{k\phi_B}j\right)e^{k\zeta_B} \qquad (25)$$

In summary, the three $S^2$ spheres and a $S^1$ unit circle $e^{k\zeta_B}$ provide a complete description of the $S^7$ two-qubit pure state. This is shown in Fig.2.

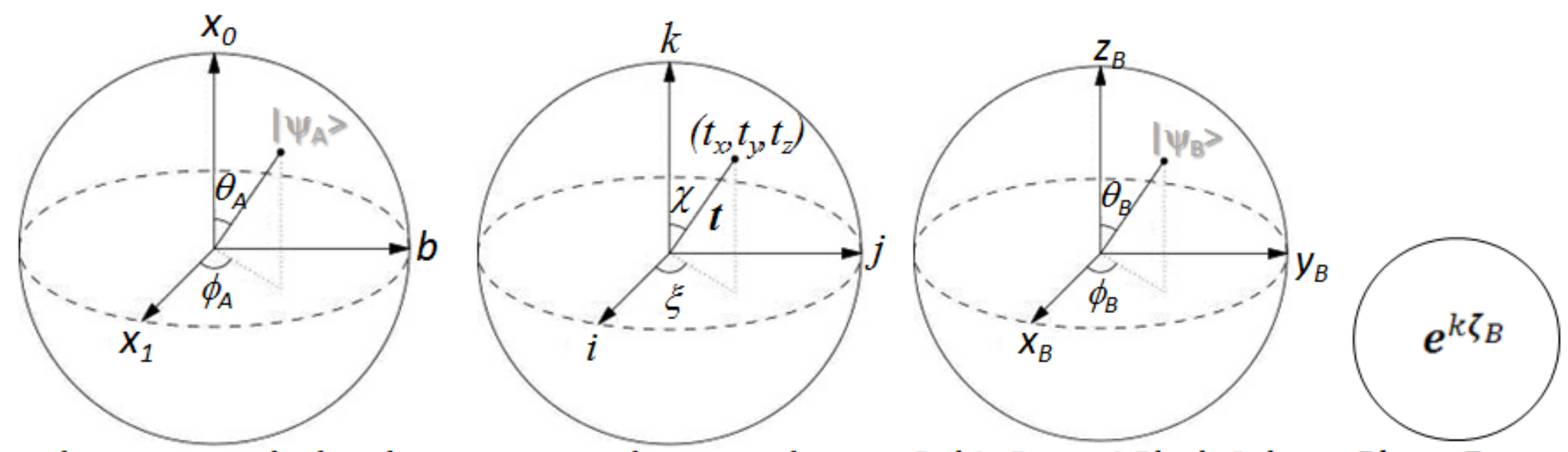

Figure 2: Three $S^2$ spheres and one $S^1$ phase factor ($e^{k\zeta_B}$). The single qubit states $|\psi_A>$ and $|\psi_B>$ do not represent a physical state but are only shown here as a placeholder for the coordinate. The entanglement sphere parameterizes the pure unit quaternion ***t***, where $\chi$ affects the degree of concurrence and $\xi$ is its phase angle (see Eq.(28)). The Bloch coordinates and the state amplitudes are related by Eq.(24). Phase factor $e^{k\zeta_B}$ affects both global phase and relative phase. For nonzero $\zeta_B$, the two phase angles need to be translated $\xi \to \xi - 2\zeta_B$ and $\phi_B \to \phi_B - 2\zeta_B$ and then, the factor $e^{k\zeta_B}$ can be regarded as the global phase factor. The concurrence is given by $c = sin\theta_A sin\phi_A sin\chi = b\ sin\chi$.

There are four important points to be mentioned here: 1) The azimuth $\xi$ of the entanglement sphere is the phase angle of concurrence. 2) For $\zeta_B \neq 0$, the phase factor $e^{k\zeta_B}$ influences the azimuth angles of both the qubit-B sphere and the entanglement sphere, and therefore care must be taken before it can be discarded as the global phase factor. 3) The qubit-A hemispheric points ($\pm b$) and the entanglement sphere antipodal points ($\pm$***t***) are identified according to $(b, t) = (-b, -t)$. This is a two-fold ambiguity. And 4) this two-qubit Bloch sphere model, Eq.(24), can not be used to find the Bloch coordinates for the separable states of the form $|\Psi_{AB}\rangle = |1\rangle_A \otimes |\psi_B\rangle$. Instead, this state may be represented as two single-qubit states. Also, if the qubit-B quasi-Bloch coordinate is $\theta_B=\pi$, then the angles $\phi_B$ and $\zeta_B$ are interchangeable. These points are discussed in more detail below.

(1) *Azimuth* $\xi$ of entanglement sphere is the phase angle of concurrence.

Writing the parameter ***t*** as

$$t = cos\chi k + sin\chi\, e^{k\xi} i \qquad (26)$$

it follows that

$$\tilde{\rho}_A = \rho_A + \frac{1}{2} c\, e^{k\left(\xi - \frac{\pi}{2}\right)} \begin{pmatrix} 0 & -j \\ j & 0 \end{pmatrix} \qquad (27)$$

where, by comparing to Eq.(10), we find

$$c e^{k\left(\xi - \frac{\pi}{2}\right)} = 2\left(\alpha\delta - \beta\gamma\right), \ |c| = concurrence, \qquad (28)$$
$$c = sin\theta_A sin\phi_A sin\chi$$

and

$$\rho_A = \frac{1}{2} \begin{pmatrix} 1 + x_0 & x_1 - x_4 k \\ x_1 + x_4 k & 1 - x_0 \end{pmatrix} \qquad (29)$$

Here, $\rho_A$ is the reduced density matrix of qubit-A, and $|c|$ is the concurrence [9]. Consider, for example, these maximally entangled states (MES) with a varying relative phase: $\frac{|00\rangle + e^{k\eta}|11\rangle}{\sqrt{2}}$ and $\frac{|01\rangle + e^{k\eta}|10\rangle}{\sqrt{2}}$ for $\eta \in [0, 2\pi)$. All of their Bloch coordinates are the same as when $\eta=0$ (which are two of the Bell states, see section IV) except for the angle $\xi$: $\xi = \pi/2 + \eta$ for the former and $\xi=3\pi/2 + \eta$ for the latter. Hence, $\xi$ parameterizes the relative phase of the entangled states. We will discuss this further in section V.

(2) *The* phase factor $e^{k\zeta_B}$ is the global phase only if it is located at the left-most end of expression. This can be seen by rewriting Eq.(24) as below:

$$\left|\tilde{\Psi}_A\right\rangle = \begin{pmatrix} \alpha + \beta j \\ \gamma + \delta j \end{pmatrix} =$$
$$e^{k\zeta_B} \begin{pmatrix} cos\frac{\theta_A}{2} \\ sin\frac{\theta_A}{2} e^{-k\zeta_B} e^{t\phi_A} e^{k\zeta_B} \end{pmatrix} \left( \cos\frac{\theta_B}{2} + \sin\frac{\theta_B}{2} e^{k(\phi_B - 2\zeta_B)} j \right) \qquad (30)$$

It is clear that the Hopf circle $S^1$($e^{k\zeta_B}$) is not simply the global phase, but it is also part of the relative phase of qubit-B. It also similarly affects the azimuth of the entanglement sphere. A unit quaternion is a spinor and a rotation operator [8, 10]. For example, $e^{-k\zeta_B} t e^{k\zeta_B}$ is another pure unit quaternion which is ***t*** rotated clockwise around the *k*-axis by angle $2\zeta_B$:

$$e^{-k\zeta_B}te^{k\zeta_B} = i\,sin\chi\cos\left(\xi-2\zeta_B\right)+j\,sin\chi\sin\left(\xi-2\zeta_B\right)+k\,cos\chi \qquad (31)$$

Hence, for $\zeta_B \neq 0$, the azimuth angles of the entanglement sphere ($\xi$) and qubit-B sphere ($\phi_B$) need be rotated clockwise by $2\zeta_B$ before ignoring the phase factor $e^{k\zeta_B}$ as the global phase.

$$\xi \to \xi\text{-}2\zeta_B;\ \text{and}\quad \phi_B \to \phi_B\text{-}2\zeta_B \qquad (32)$$

Note that this same phase factor must be factored out of the state amplitudes $\alpha, \beta, \gamma, \delta$ for Eqs.(24) and (30) to remain consistent after the global phase factor is discarded.

(3) *The* $\pm b$ hemispheric points of qubit-A sphere and the $\pm t$ antipodal points of the entanglement sphere are identified according to: $(b,t)=(-b,-t)$; or $(b,-t)=(-b,t)$. This introduces a two-fold ambiguity in the combined coordinates of the qubit-A sphere and the entanglement sphere. This is because only the ***bt*** product appears in the base parameters:

$$\tilde{\rho}_A = \frac{1}{2}\begin{pmatrix} 1+x_0 & x_1-bt \\ x_1+bt & 1-x_0 \end{pmatrix} \quad \text{and}$$

$$\left|\tilde{\Psi}_A\right\rangle = \frac{1}{\sqrt{2\left(1+x_0\right)}}\begin{pmatrix} 1+x_0 \\ x_1+bt \end{pmatrix} q_B \qquad (33)$$

The Bloch sphere coordinates with the same ***bt*** product (while all other coordinates are being the same) correspond to the same state. On a different note, Eq.(33) can be used as a shortcut to Eq.(24) when determining the Bloch sphere coordinates. The column vector in the quasi state can be taken straight from the first column of the quasi density matrix. One can determine the values of $x_0$, $x_1$, $b$ and $t$ by comparing $\tilde{\rho}_A$ of Eq.(33) with Eq.(10), and noting that ***t*** is a pure-imaginary unit quaternion. A detailed step-by-step parameter determination procedure of the seven angle parameters is given in the next section (section IV).

(4) *This* Bloch sphere parameterization of Eq.(24) is not valid for a separable state given by $\left|\Psi_{AB}\right\rangle = \left|1\right\rangle_A \otimes \left|\psi_B\right\rangle$. For this state, qubit-A is at the South pole, $x_0$= -1 or $\theta_A=\pi$. In this case, the qubit-B can be plotted on its Bloch sphere using the single qubit state $\left|\psi_B\right\rangle$. This separable state appears to be the only exception to our model, Eq.(24). Another minor issue occurs when the qubit-B is at the South pole: $z_B$= -1 or $\theta_B=\pi$. In this case, the terms, $e^{k\phi_B}$ and $e^{k\zeta_B}$, are interchangeable in $q_B$ with the appropriate change of sign. The final state, however, is consistent after factoring out the global phase factor.

# IV. PROCEDURE TO OBTAIN ($\theta_A$, $\phi_A$), ($\chi$, $\xi$), ($\theta_B$, $\phi_B$), $\zeta_B$ FROM $\alpha,\beta,\gamma,\delta$, AND EXAMPLES

## A. Parameter determination procedure

It is straightforward to obtain the amplitudes $\alpha$, $\beta$, $\gamma$, and $\delta$ from the seven angle parameters (we used imaginary unit $i$ in the formulas of Eq.(34)):

$$\alpha = cos\frac{\theta_A}{2}cos\frac{\theta_B}{2}e^{i\zeta_B};\ \beta = cos\frac{\theta_A}{2}sin\frac{\theta_B}{2}e^{i(\phi_B-\zeta_B)};$$

$$\gamma = sin\frac{\theta_A}{2}\begin{bmatrix}\left(cos\phi_A + i\,sin\phi_A cos\chi\right)cos\frac{\theta_B}{2}+ \\ i\,sin\phi_A sin\chi\sin\frac{\theta_B}{2}e^{i(\xi-\phi_B)}\end{bmatrix}e^{i\zeta_B}; \qquad (34)$$

$$\delta = sin\frac{\theta_A}{2}\begin{bmatrix}\left(cos\phi_A + i\,sin\phi_A cos\chi\right)sin\frac{\theta_B}{2}- \\ i\,sin\phi_A sin\chi\cos\frac{\theta_B}{2}e^{i(\xi-\phi_B)}\end{bmatrix}e^{i(\phi_B-\zeta_B)}$$

In these expressions, it is clear that factoring out the global phase $e^{i\zeta_B}$ requires the azimuth angles $\xi$ and $\phi_B$ to be shifted accordingly: $\xi \to \xi\text{-}2\zeta_B$ and $\phi_B \to \phi_B\text{-}2\zeta_B$. Note here that since no quaternion is involved in Eq.(34), the imaginary unit $k$ was replaced by $i$. We present in this section a streamlined procedure on how to obtain the Bloch coordinates (the seven angles) from the complex amplitudes $\alpha, \beta, \gamma$, and $\delta$.

The four angle parameters ($\theta_A$, $\phi_A$, $\chi$, $\xi$) and their Cartesian coordinates ($x_0$, $x_1$, $x_2$, $x_3$, $x_4$) of the Hopf base are determined from the quasi density matrix $\tilde{\rho}_A$ which is given below in Eq.(39). Once the base parameters have been found, the three parameters ($\theta_B$, $\phi_B$, $\zeta_B$) of the Hopf fiber $q_B$ are found from the quasi state $\left|\tilde{\Psi}_A\right\rangle$, Eqs.(35) and (25). From Eqs.(24) and (25),

$$\left|\tilde{\Psi}_A\right\rangle = \begin{pmatrix}\alpha+\beta j \\ \gamma+\delta j\end{pmatrix} = \begin{pmatrix} cos\frac{\theta_A}{2} \\ sin\frac{\theta_A}{2}e^{t\phi_A}\end{pmatrix} q_B \qquad (35)$$

the quasi density matrix is

$$\tilde{\rho}_A \equiv \left|\tilde{\Psi}_A\right\rangle\left\langle\tilde{\Psi}_A\right| = \begin{pmatrix} cos\frac{\theta_A}{2} \\ sin\frac{\theta_A}{2}e^{t\phi_A}\end{pmatrix} q_B\bar{q}_B\left(cos\frac{\theta_A}{2}, sin\frac{\theta_A}{2}e^{-t\phi_A}\right)$$

$$= \frac{1}{2}\left(\boldsymbol{I} + \boldsymbol{n}_A \cdot \boldsymbol{\sigma}(t)\right) \qquad (36)$$

Where,

$$q_B \bar{q}_B = 1,\ \boldsymbol{\sigma}(t) = \left(\sigma_x, \sigma_y(t), \sigma_z\right),$$
$$\boldsymbol{n}_A = \left(sin\theta_A cos\phi_A, sin\theta_A sin\phi_A, cos\theta_A\right) = \left(x_1, b, x_0\right)$$
$$t = sin\chi cos\xi\, i + sin\chi sin\xi\, j + cos\chi k\,;$$
$$\theta_A, \theta_B,\ \chi \in [0, \pi); \phi_A, \phi_B, \zeta_B, \xi \in [0, 2\pi) \quad (37)$$

The Pauli matrices are defined as

$$\sigma_x = \begin{pmatrix} 0 & 1 \\ 1 & 0 \end{pmatrix}, \sigma_y(t) = \begin{pmatrix} 0 & -t \\ t & 0 \end{pmatrix}, \sigma_z = \begin{pmatrix} 1 & 0 \\ 0 & -1 \end{pmatrix} \quad (38)$$

The Hopf fiber $q_B$ (parameters: $\theta_B$, $\phi_B$, $\zeta_B$) is completely eliminated from $\tilde{\rho}_A$. Hence, $\tilde{\rho}_A$ depends only on $\boldsymbol{t}$ and the qubit-A parameters. Here, $\boldsymbol{t}$ is the imaginary unit in the Pauli matrix $\sigma_y(t)$. The quasi-density matrix $\tilde{\rho}_A$ is written in full as

$$\tilde{\rho}_A = \frac{1}{2}\begin{pmatrix} 1+x_0 & x_1 - bt \\ x_1 + bt & 1 - x_0 \end{pmatrix} =$$
$$\frac{1}{2}\begin{pmatrix} 1+x_0 & \left(x_1 - x_4 k\right) - \left(x_3 - x_2 k\right) j \\ \left(x_1 + x_4 k\right) + \left(x_3 - x_2 k\right) j & 1 - x_0 \end{pmatrix} \quad (39)$$
$$= \begin{pmatrix} |\alpha|^2 + |\beta|^2 & \left(\alpha\bar{\gamma} + \beta\bar{\delta}\right) - \left(\alpha\delta - \beta\gamma\right) j \\ \left(\bar{\alpha}\gamma + \bar{\beta}\delta\right) + \left(\alpha\delta - \beta\gamma\right) j & |\gamma|^2 + |\delta|^2 \end{pmatrix}$$

where,

$$x_0 = cos\theta_A,\ x_1 = sin\theta_A cos\phi_A,\ b = sin\theta_A sin\phi_A,\ c = b\, sin\chi,$$
$$x_2 = c\, cos\xi\ ,\ x_3 = c\, sin\xi,\ x_4 = b\, cos\chi, bt = x_2 i + x_3 j + x_4 k$$

All angle parameters of the $S^4$ Hopf base, ($\theta_A$, $\phi_A$) and ($\chi$, $\xi$), and the equivalent Cartesian coordinates ($x_0$, $x_1$, $x_2$, $x_3$, $x_4$), are obtained. The procedure is outlined step by step below:

1. *$S^4$ Hopf base parameters:($\theta_A$, $\phi_A$),($\chi$, $\xi$)⇔($x_0$, $x_1$, $x_2$, $x_3$, $x_4$)*

**step-1)** $$x_0 = cos\theta_A = |\alpha|^2 + |\beta|^2 - |\gamma|^2 - |\delta|^2 \Rightarrow \theta_A$$
$$sin\theta_A = \sqrt{1 - x_0{}^2},\ cos\frac{\theta_A}{2} = \sqrt{\frac{1+x_0}{2}},\ sin\frac{\theta_A}{2} = \sqrt{\frac{1-x_0}{2}}$$

**step-2)** $$x_1 + kx_4 = 2\left(\bar{\alpha}\gamma + \bar{\beta}\delta\right) \Rightarrow \phi_A, \chi$$
$$x_1 = sin\theta_A cos\phi_A \rightarrow \phi_A\left(two\ possible\ values\right)$$
$$\rightarrow b = sin\theta_A sin\phi_A$$
$$x_4 = b\, cos\chi \rightarrow \chi \rightarrow c = b\, sin\chi$$

**step-3)** $$x_3 - kx_2 = 2\left(\alpha\delta - \beta\gamma\right) \Rightarrow \xi$$
$$x_2 = c\, cos\xi,\ x_3 = c\, sin\xi\ \rightarrow \xi \rightarrow$$
$$t = i\, sin\chi cos\xi + j\, sin\chi sin\xi + k\, cos\chi$$

2. *$S^3$ Hopf fiber parameters: ($\theta_B$, $\phi_B$) and $\zeta_B$*

Once the base parameters are determined, the parameters, ($\theta_B$, $\phi_B$) and $\zeta_B$, are obtained from the quasi state $|\tilde{\Psi}_A\rangle$.

**step-4)** $$|\tilde{\Psi}_A\rangle = \begin{pmatrix} cos\frac{\theta_A}{2} \\ sin\frac{\theta_A}{2} e^{t\phi_A} \end{pmatrix} q_B = \begin{pmatrix} \alpha + \beta j \\ \gamma + \delta j \end{pmatrix} \Rightarrow q_B \Rightarrow \theta_B, \phi_B, \zeta_B$$

Solving for $q_B$, we get

$$q_B = cos\frac{\theta_A}{2}\left(\alpha + \beta j\right) + sin\frac{\theta_A}{2} e^{-t\phi_A}\left(\gamma + \delta j\right)$$

The qubit-B parameters are found by equating this $q_B$ to

$$q_B = \left(\cos\frac{\theta_B}{2} + \sin\frac{\theta_B}{2} e^{k\phi_B} j\right) e^{k\zeta_B} \rightarrow \theta_B, \phi_B, \zeta_B$$

Through these four steps, the seven angle parameters are obtained from the four complex amplitudes of the state. At this point, if $\zeta_B \neq 0$, then $2\zeta_B$ may be subtracted from $\xi$ and $\phi_B$ (where they become the new $\xi$ and $\phi_B$) while the global phase factor $e^{k\zeta_B}$ is factored out of the complex amplitudes $\alpha, \beta, \gamma, \delta$. Then the global phase can be reset (that is, set $\zeta_B$=0) or ignored. All these four steps can be done quickly using the shortcut Eq.(33).

### B. Examples of two-qubit states on Bloch sphere

#### 1. *Maximally entangled states, Bell states*

As a first example, we plot the four maximally entangled states (MES), namely the Bell states defined as:

$$|\beta_{00}\rangle = \frac{|00\rangle + |11\rangle}{\sqrt{2}}, |\beta_{01}\rangle = \frac{|01\rangle + |10\rangle}{\sqrt{2}}, \quad (40)$$
$$|\beta_{10}\rangle = \frac{|00\rangle - |11\rangle}{\sqrt{2}}, |\beta_{11}\rangle = \frac{|01\rangle - |10\rangle}{\sqrt{2}}$$

In the qubit-A basis, converting qubit-B basis states to quaternion units, we get the following quasi states:

$$|\tilde{\beta}_{00}\rangle = \frac{1}{\sqrt{2}}\begin{pmatrix} 1 \\ j \end{pmatrix}, |\tilde{\beta}_{01}\rangle = \frac{1}{\sqrt{2}}\begin{pmatrix} j \\ 1 \end{pmatrix}, \quad (41)$$
$$|\tilde{\beta}_{10}\rangle = \frac{1}{\sqrt{2}}\begin{pmatrix} 1 \\ -j \end{pmatrix}, |\tilde{\beta}_{11}\rangle = \frac{1}{\sqrt{2}}\begin{pmatrix} j \\ -1 \end{pmatrix}$$

The quasi density matrices are:

$$\tilde{\rho}_{00} = \frac{1}{2}\begin{pmatrix} 1 & -j \\ j & 1 \end{pmatrix}, \tilde{\rho}_{01} = \frac{1}{2}\begin{pmatrix} 1 & j \\ -j & 1 \end{pmatrix}, \quad (42)$$
$$\tilde{\rho}_{10} = \frac{1}{2}\begin{pmatrix} 1 & j \\ -j & 1 \end{pmatrix}, \tilde{\rho}_{11} = \frac{1}{2}\begin{pmatrix} 1 & -j \\ j & 1 \end{pmatrix}$$

For this example we use the shortcut Eq.(33). From the density matrix we find:

$\tilde{\rho}_{00}$ : $x_0$=0, $x_1$=0, b=1, t=j

$\tilde{\rho}_{01}$ : $x_0$=0, $x_1$=0, b=1, t=-j

$\tilde{\rho}_{10}$ : $x_0$=0, $x_1$=0, b=1, t=-j

$\tilde{\rho}_{11}$ : $x_0$=0, $x_1$=0, b=1, t=j

From the quasi state in Eq.(33), equating it to Eq.(41), we find $q_B$, which we then equate to Eq.(25), to find

$|\tilde{\beta}_{00}\rangle$: $q_B$=1 → $\theta_B$=0, $\zeta_B$=0

$|\tilde{\beta}_{01}\rangle$: $q_B$=j → $\theta_B$=π, $\phi_B$=0, $\zeta_B$=0

$|\tilde{\beta}_{10}\rangle$: $q_B$=1 → $\theta_B$=0, $\zeta_B$=0

$|\tilde{\beta}_{11}\rangle$: $q_B$=j → $\theta_B$=π, $\phi_B$=0, $\zeta_B$=0

The Bloch coordinates and Bloch spheres are summarized in the Table I and Table II, respectively.

Table I The Bloch coordinates of maximally entangled Bell states. Since (b, t) and (-b, -t) give the same state, the coordinate for $|\beta_{cd}\rangle$ alternate to the table is *cd* = *($x_1$, b, $x_0$)* & *($t_x$, $t_y$, $t_z$)*: 00 = (0,-1,0) & (0,-1,0); 01 = (0,-1,0) & (0, 1, 0); 10 = (0,-1,0) & (0, 1, 0); 11 = (0,-1,0) & (0, -1, 0).

| Bloch sphere Coordinates | $\|\beta_{00}\rangle$ | $\|\beta_{01}\rangle$ | $\|\beta_{10}\rangle$ | $\|\beta_{10}\rangle$ |
|---|---|---|---|---|
| ($x_0$, b, $x_1$) | (0, 1, 0) | (0, 1, 0) | (0, 1, 0) | (0, 1, 0) |
| ($t_x$, $t_y$, $t_z$) | (0, 1, 0) | (0, -1, 0) | (0, -1, 0) | (0, 1, 0) |
| ($x_B$, $y_B$, $z_B$) | (0, 0, 1) | (0, 0, -1) | (0, 0, 1) | (0, 0, -1) |
| $\zeta_B$ | 0 | 0 | 0 | 0 |

Table II Bell State Bloch Spheres. For all cases, $\zeta_B$=0. The first three Bloch sphere rows are for the Bell states. In the first two rows, filled (red) and hollow (blue) dots are the alternate coordinates (b,t) and (-b,-t), respectively. The last row is the qubit-A Bloch sphere obtained by applying CNOT gate to the Bell states. The CNOT gate did not affect the qubit-B Bloch coordinate.

| | $\|\beta_{00}\rangle$ | $\|\beta_{01}\rangle$ | $\|\beta_{10}\rangle$ | $\|\beta_{11}\rangle$ |
|---|---|---|---|---|
| Qubit-A Bloch Sphere | | | | |
| Entanglement Sphere | | | | |
| Qubit-B Sphere before and after CNOT | | | | |
| CNOT | CNOT$\|\beta_{00}\rangle$ | CNOT$\|\beta_{01}\rangle$ | CNOT$\|\beta_{10}\rangle$ | CNOT$\|\beta_{11}\rangle$ |
| Qubit-A after CNOT | | | | |

Applying the CNOT gate to the Bell state, a product state results. The qubit-B state remained unchanged, while qubit-A coordinate changed to $x_1 = \pm 1$, $b = 0$ (which gives zero concurrence, $c = b\ sin\chi = 0$). The entanglement sphere can be ignored for any state with $b$=0. The product state can be easily written down from the Bloch coordinates. For the entangled states, the states can be recovered from the Bloch coordinates using Eq.(24).

*2. Separable states*

For separable states the concurrence is zero, which is satisfied if $b = sin\theta_A sin\phi_A = 0$ or if $sin\chi = 0$: the former condition (b=0) is satisfied if the qubit-A is on the great circle on the $x_0x_1$-plane ($b=0$). In this case, the entanglement sphere is undefined because ***t*** appears in the quasi-state and quasi-density matrix only as a product with $b$ (see Eq.(33)). In the second case ($sin\chi=0$), the entanglement sphere coordinate is either at the North pole ($\chi=0$) or at the South pole ($\chi=\pi$). In both cases, since the concurrence is zero ($c=0$), the angle ξ is undefined (see **step-3**). From the two-fold ambiguity in $(b, t)$, choosing $t=k$ (i.e., χ=0, North pole), the qubit-A Bloch sphere coordinate agrees with the single qubit state of the original product state.

For a separable state where ***t*** equals $k$, the Hopf base $S^4$ is reduced to $S^2$:

$$\underbrace{\begin{pmatrix} cos\dfrac{\theta_A}{2} \\ sin\dfrac{\theta_A}{2}e^{t\phi_A} \end{pmatrix}}_{S^4} \rightarrow \underbrace{\begin{pmatrix} cos\dfrac{\theta_A}{2} \\ sin\dfrac{\theta_A}{2}e^{k\phi_A} \end{pmatrix}}_{S^2} \tag{43}$$

In this case, the quasi state

$$\left|\tilde{\Psi}_A\right\rangle = e^{k\zeta_B}\begin{pmatrix} cos\dfrac{\theta_A}{2} \\ sin\dfrac{\theta_A}{2}e^{k\phi_A} \end{pmatrix}\left(\cos\frac{\theta_B}{2} + \sin\frac{\theta_B}{2}e^{k(\phi_B - 2\zeta_B)}j\right) \tag{44}$$

corresponds to the composite state:

$$\left|\Psi_{AB}\right\rangle = e^{k\zeta_B}\begin{pmatrix} cos\dfrac{\theta_A}{2} \\ sin\dfrac{\theta_A}{2}e^{k\phi_A} \end{pmatrix} \otimes \begin{pmatrix} cos\dfrac{\theta_B}{2} \\ \sin\dfrac{\theta_B}{2}e^{k(\phi_B - 2\zeta_B)} \end{pmatrix} \equiv \left|\psi_A\right\rangle \otimes \left|\psi_B\right\rangle \tag{45}$$

The global phase factor $e^{k\zeta_B}$ may be ignored (*i.e.*, set to unity) after the azimuth of qubit-B Bloch sphere has been translated: $\phi_B \rightarrow \phi_B - 2\zeta_B$. It is clear that when $t = k$ (or $-k$), the entire coordinate of qubit-A Bloch sphere is valid for the single qubit state. The separable state Bloch spheres are shown in Fig.3. Some specific examples are shown in the Table III.

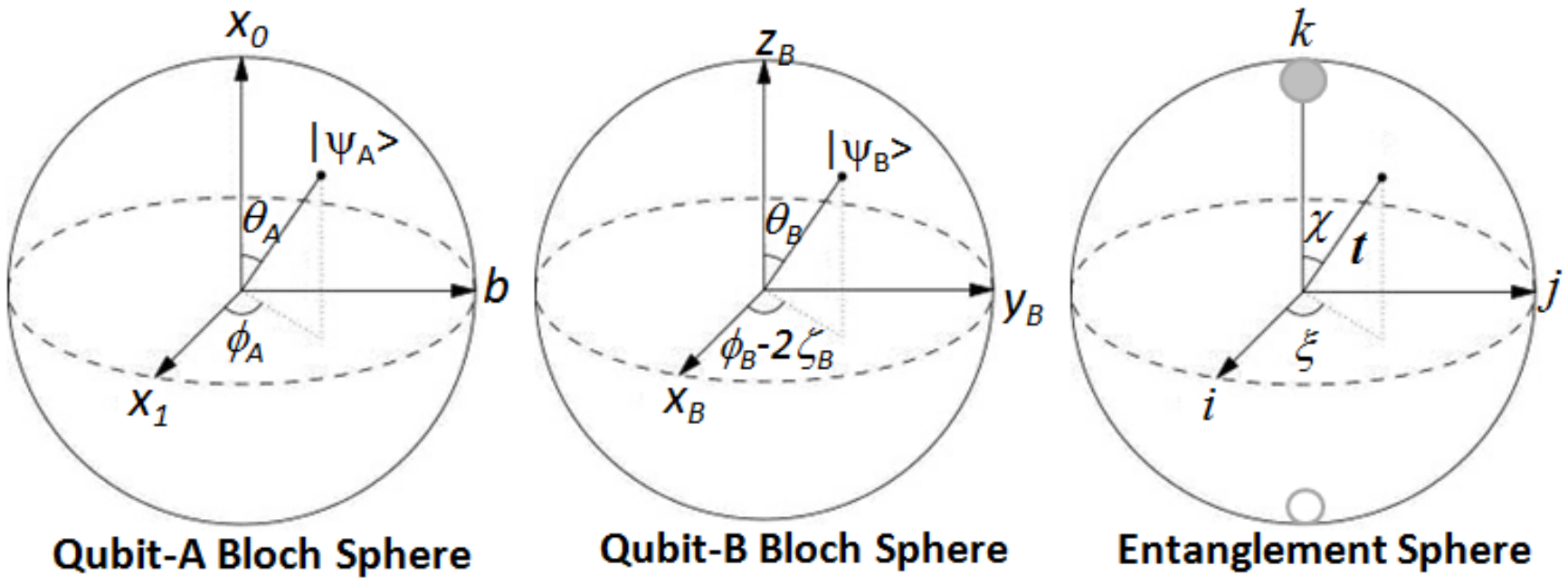

Figure 3: The separable state Bloch spheres of Eq.(45). The entanglement sphere is necessary when the qubit-A Bloch coordinate is at $b \neq 0$, and in which case, the entanglement sphere coordinate must be either the North pole (filled gray dot) or the South pole (open gray dot).

Table III Separable state examples: $\left|\Psi_{AB}\right\rangle = \left|\psi_A\right\rangle \otimes \left|\psi_B\right\rangle$. In all cases, $\zeta_B = 0$. In the coordinates of qubit-A sphere and entanglement sphere, the two pairs of corresponding shape (or color) coordinates give the same ***bt*** product in Eq.(33). The North pole (filled red dot) of entanglement sphere (i.e., $t = k$) gives qubit-A coordinate which agrees with the single-qubit state $|\psi_A>$.

| $\lvert\psi_A\rangle$ | $\lvert\psi_B\rangle$ | Qubit-A Sphere | Entanglement Sphere | Qubit-B Sphere |
|---|---|---|---|---|
| $\frac{\lvert 0\rangle + \lvert 1\rangle}{\sqrt{2}}$ | $\frac{\lvert 0\rangle - k\lvert 1\rangle}{\sqrt{2}}$ | | UNDEFINED | |
| $\frac{\lvert 0\rangle - k\lvert 1\rangle}{\sqrt{2}}$ | $\frac{\lvert 0\rangle + k\lvert 1\rangle}{\sqrt{2}}$ | | | |
| $\frac{\lvert 0\rangle + k\lvert 1\rangle}{\sqrt{2}}$ | $\frac{\lvert 0\rangle - \lvert 1\rangle}{\sqrt{2}}$ | | | |
| $\frac{\lvert 0\rangle + e^{-k\frac{\pi}{4}}\lvert 1\rangle}{\sqrt{2}}$ | $\lvert 1\rangle$ | | | |

### C. Trajectory of CNOT, CZ and SWAP gates

We plot the Bloch sphere trajectory of two-qubit gates. Let us consider a controlled-unitary, C(U), gate in two steps. Here, U is a single-qubit unitary which can be written as

$$U = e^{k\eta} R_{\boldsymbol{n}}(\omega) = e^{k\eta} e^{-k\frac{\omega}{2}\boldsymbol{n}\cdot\boldsymbol{\sigma}} \qquad (46)$$

where η is a phase angle, $R_{\boldsymbol{n}}(\omega)$ is the rotation operator that rotates a single-qubit state by an angle ω around the axis $\boldsymbol{n}$, and $\boldsymbol{\sigma}$ is the Pauli matrix, $\boldsymbol{\sigma} = (\sigma_x, \sigma_y(k), \sigma_z)$. The Pauli X and Z gates can be written as

$$X = e^{k\frac{\pi}{2}} R_x(\pi) \text{ to be used in CNOT and SWAP}$$

$$Z = e^{k\frac{\pi}{2}} R_z(\pi) \text{ to be used in CZ}$$

We take a two step process: First, the phase angle η increases from 0 to π/2, η = 0→π/2 (marked '1' in the figure) while the rotation angle is fixed at zero, ω=0. Then, the phase is fixed at π/2, η=π/2, and the rotation angle increases from ω=0 to ω =π around a given axis (marked '2' in the figure).

$$'1' = e^{k\eta} R_{\boldsymbol{n}}(0) = e^{k\eta}, \eta = 0 \rightarrow \frac{\pi}{2};$$
$$'2' = e^{k\frac{\pi}{2}} R_{\boldsymbol{n}}(\omega) = kR_{\boldsymbol{n}}(\omega), \omega = 0 \rightarrow \pi \qquad (47)$$

We consider the trajectories of CNOT, CZ and SWAP gates following these two steps.

#### *1. CNOT and CZ gates*

Let us take $\lvert\beta_{10}\rangle = \frac{\lvert 00\rangle - \lvert 11\rangle}{\sqrt{2}}$ as an example. The CNOT gate and CZ gate trajectories are shown in Figures 4 and 5, respectively. For the CZ gate in Figure 5, the entire process amounts to a change in the nonlocal phase angle ξ by π (from ξ=3π/2 to π/2 counterclockwise).

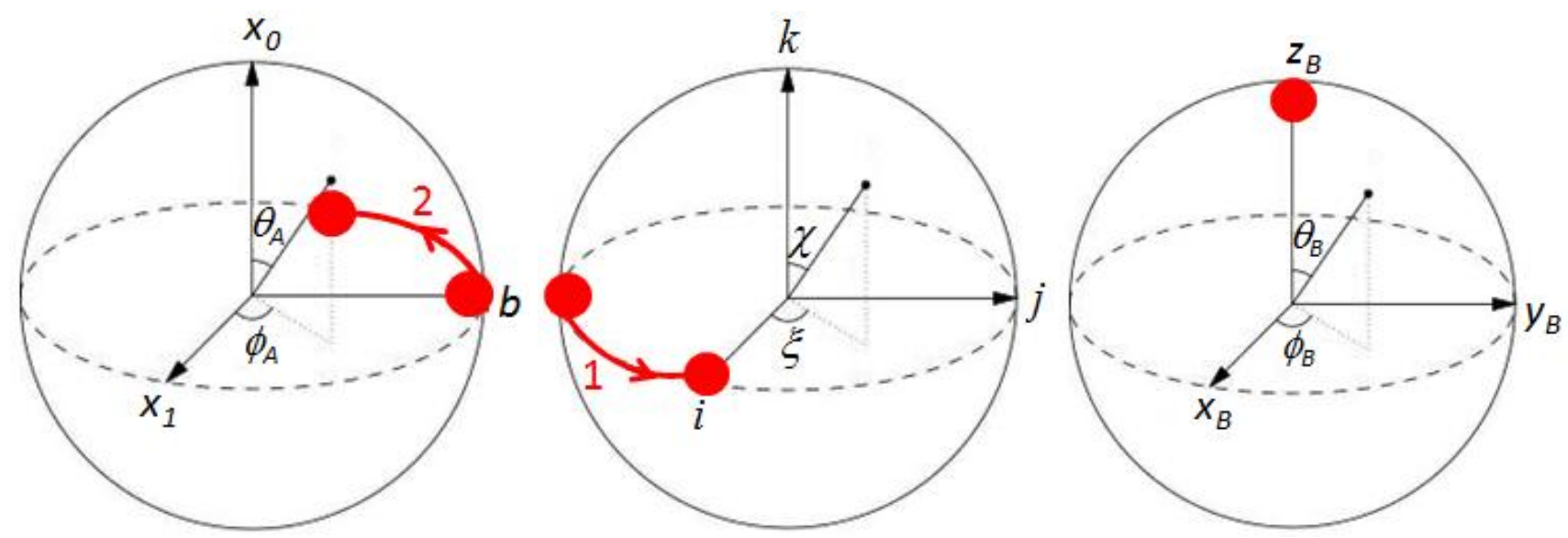

Figure 4: Evolution of CNOT gate on the Bell state $|\beta_{10}\rangle$: $CNOT|\beta_{10}\rangle = \frac{|0\rangle - |1\rangle}{\sqrt{2}} \otimes |0\rangle$. The alternate *(b, t) =(-b,-t)* trajectory is not shown. $\zeta_B$=0. See Eq.(47).

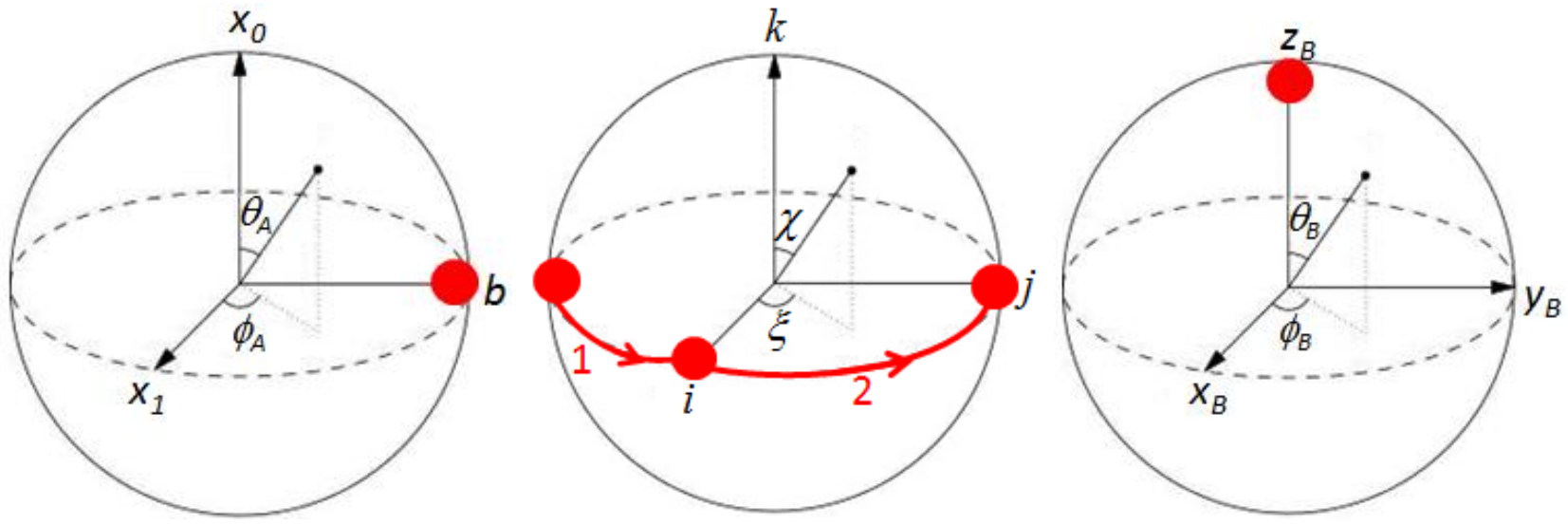

Figure 5: Evolution of the CZ gate on the Bell state $|\beta_{10}\rangle$: CZ$|\beta_{10}\rangle$=$|\beta_{00}\rangle$. The alternate *(b, t) =(-b,-t)* is not shown. $\zeta_B$=0. See Eq.(47).

### 2. *SWAP gate*

We consider the SWAP operation as

$$SWAP\begin{pmatrix}\alpha\\ \beta\\ \gamma\\ \delta\end{pmatrix} = \begin{pmatrix}\alpha\\ X\begin{pmatrix}\beta\\ \gamma\end{pmatrix}\\ \delta\end{pmatrix} = \begin{pmatrix}\alpha\\ e^{k\frac{\pi}{2}}R_x(\pi)\begin{pmatrix}\beta\\ \gamma\end{pmatrix}\\ \delta\end{pmatrix} = \begin{pmatrix}\alpha\\ e^{k\eta}R_x(\omega)\begin{pmatrix}\beta\\ \gamma\end{pmatrix}\\ \delta\end{pmatrix}_{\substack{\eta=0\to\frac{\pi}{2},\\ \omega=0\to\pi}}$$

Similar to the CNOT and CZ gates above, we first let the phase evolve $\eta = 0 \to \frac{\pi}{2}$ which is marked as '1' in the figure, followed by an X-rotation $\omega = 0 \to \pi$ which is marked '2' in the figure. As an example, in Fig.6 we show the SWAP gate operating on a separable state $|\Psi_{AB}\rangle = \frac{|01\rangle + k|11\rangle}{\sqrt{2}}$. The resulting state is another separable state $\frac{|10\rangle + k|11\rangle}{\sqrt{2}}$.

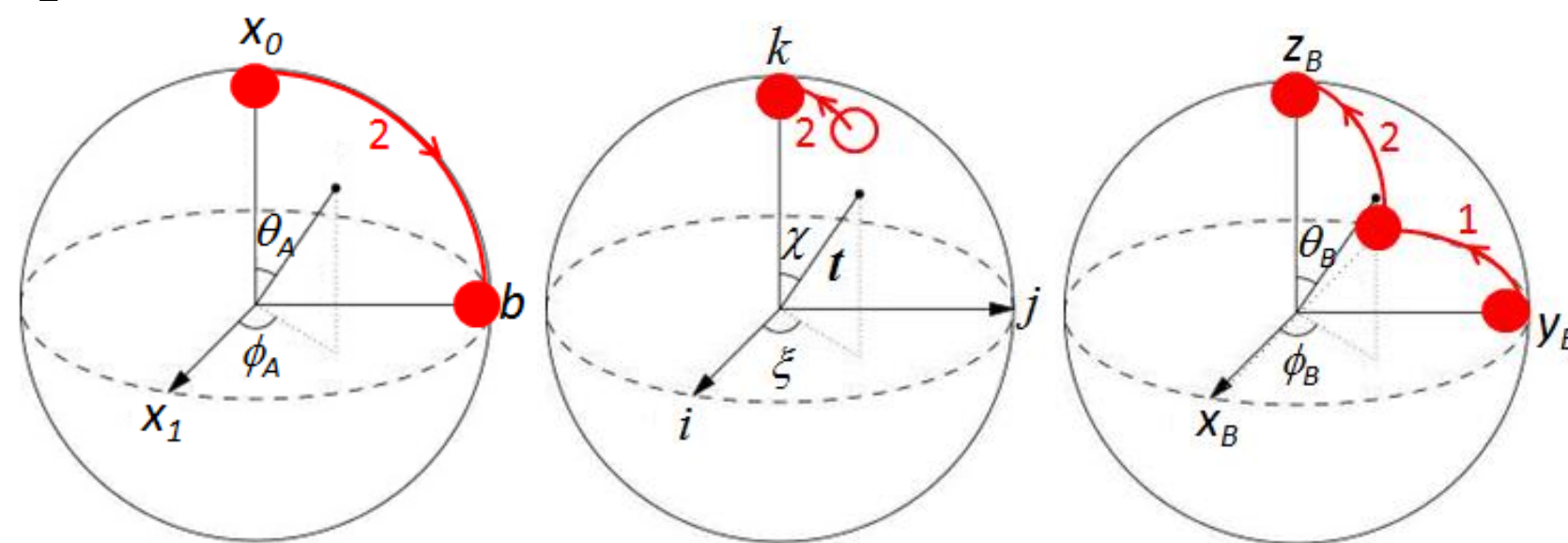

Figure 6: The action of a SWAP gate on $\frac{|00\rangle + k|01\rangle}{\sqrt{2}} = |0\rangle \otimes \frac{|0\rangle + k|1\rangle}{\sqrt{2}}$: $SWAP\frac{|00\rangle + k|01\rangle}{\sqrt{2}} = \frac{|00\rangle + k|10\rangle}{\sqrt{2}} = \frac{|0\rangle + k|1\rangle}{\sqrt{2}} \otimes |0\rangle$. The hollow red circle in the entanglement sphere is the initial state at $\eta$=$\pi$/2, $\omega$=0. $\zeta_B$=0.

## V. DISCUSSION

The two quasi-Bloch spheres seem to be a natural choice for their name and coordinates because for separable states, these spheres reduce to the respective single qubit Bloch sphere (with the azimuth of qubit-B Bloch sphere adjusted to $\phi_B$-$2\zeta_B$). The phase angle $\zeta_B$ is not simply the global phase, but it also affects the azimuth angles $\phi_B$ and $\xi$. After properly adjusting these two azimuth angles (that is, after subtracting $2\zeta_B$ from each and replacing the azimuth with these new values, Eq.(32)), the phase factor $exp(k\zeta_B)$ can be considered a pure global phase factor and ignored.

The entanglement sphere and its coordinates deserve more discussion. This sphere seems to capture the nonlocal parameters. The maximally entangled states exist on the equator of this sphere ($\chi = \pi/2$) together with the possible two coordinates $b=\pm 1$ (East and West) on the equator of qubit-A Bloch sphere. If $\chi = 0$ or $\pi$, then the state is separable and the Bloch spheres may be considered the single-qubit Bloch sphere of each qubit. A product state can also happen if the qubit-A coordinate is on the great circle corresponding to $b=0$ (prime meridian of the qubit-A sphere).

The azimuth $\xi$ of the entanglement sphere deserves much more discussion. This appears to parameterize a 'nonlocal relative phase' angle of the composite state. As was mentioned in section III (see Eqs.(27) and (28) and discussion there) this angle is the phase angle of concurrence. It was mentioned that this angle is equal to the relative phase angle of MES states offset by some fixed constant. Furthermore, the azimuth angles $\phi_A$ and $\phi_B$ seem to parameterize the 'local' relative phase angle in the sense that they capture the relative phase angle between the 'separable pairs' of the basis components in the composite state. To be more specific, let us consider the following state:

$$\left|\Psi_{AB}\right\rangle = e^{k\eta}\left[ae^{-k\varphi_1}\left|00\right\rangle + be^{-k\varphi_2}\left|01\right\rangle + ce^{k\varphi_2}\left|10\right\rangle + de^{k\varphi_1}\left|11\right\rangle\right] \tag{48}$$

where $a$, $b$, $c$, $d$ are a non-negative real number, normalized; and $\eta$, $\varphi_1$ and $\varphi_2$ are real numbers. This is a state with a definite phase factor for the concurrence (where, $\alpha\delta - \beta\gamma = (ad-bc)e^{k2\eta}$) while the component basis states can vary with a relative phase from each other. Suppose we consider a pair of component states at a time. The $a$-$b$ pair is for a product state with |0> for qubit-A with qubit-B having a relative phase factor $exp[k(\varphi_1-\varphi_2)]$ between its two basis states. For the $c$-$d$ pair, qubit-A is in |1> and qubit-B has a relative phase factor $exp[-k(\varphi_1-\varphi_2)]$ between its two basis states. Therefore, the qubit-B quasi-Bloch sphere would have its $\phi_B$ coordinate depend on $\varphi_1$-$\varphi_2$. The same argument applies to the $a$-$c$ and $b$-$d$ pairs, leading us to expect the qubit-A Bloch coordinate $\phi_A$ to depend on $\varphi_1+\varphi_2$. Moreover, the $a$-$d$ and the $b$-$c$ pairs are entangled pairs with a relative phase factor $exp(2k\varphi_1)$ and $exp(2k\varphi_2)$, respectively. In section III, it was mentioned for the two MES states that $\xi = 2\varphi_1+\pi/2$ for $a=1/\sqrt{2}=d$ $(b=0=c)$, and $\xi=2\varphi_2+3\pi/2$ for $b=1/\sqrt{2}=c$ $(a=0=d)$. Another example may be $a=b=d=1/\sqrt{3}$, $c=0$, which is a partially entangled state (concurrence = 2/3). For this state we obtain $\xi=2\varphi_1+\pi/2$, $\phi_B=\varphi_1$-$\varphi_2$, and $cos\phi_A = (1/\sqrt{2})cos(\varphi_1+\varphi_2)$. For a separable state example, we find $\phi_B=\varphi_1$-$\varphi_2$ for $a=1/\sqrt{2}=b$ $(c=0=d)$; $\phi_A=\varphi_1+\varphi_2$ for $b=1/\sqrt{2}=d$ $(a=0=c)$; and $\phi_A=\varphi_1+\varphi_2$ and $\phi_B=\varphi_1$-$\varphi_2$ for $a=b=c=d=1/2$. In all cases, any non-zero global phase angle $\zeta_B$ had been factored out to the left end of expression as in Eq.(30) with the angles $\xi$ and $\phi_B$ properly shifted. These examples suggest that $\xi$ is a 'nonlocal' relative phase angle of the composite state while $\phi_A$ and $\phi_B$ are the 'local' relative phase angles. However, this interpretation needs more investigation.

Let us discuss the single-qubit Bloch sphere (or ball) of qubit-A after a partial trace over qubit-B. Once qubit-B is traced out, the quasi-state vector $\left|\tilde{\Psi}_A\right\rangle$ is no longer a well defined quantity. The partial trace over qubit-B maps the two-quit density matrix $\rho_{AB}=|\Psi_{AB}><\Psi_{AB}|$ to the reduced density matrix $\rho_A$. Similarly, let us define the partial trace over qubit-B as a map on the quasi-density matrix $\tilde{\rho}_A$, $tr_B$: $\tilde{\rho}_A \rightarrow \rho_A$.

$$tr_B:\ \tilde{\rho}_A = \frac{1}{2}\begin{pmatrix} 1+x_0 & x_1-bt \\ x_1+bt & 1-x_0 \end{pmatrix} \rightarrow \qquad \rho_A = \frac{1}{2}\begin{pmatrix} 1+x_0 & x_1-x_4k \\ x_1+x_4k & 1-x_0 \end{pmatrix} \tag{49}$$

Let us note first that $\tilde{\rho}_A$ has the properties of a pure-state density matrix: $\tilde{\rho}_A^{\,2} = \tilde{\rho}_A$, tr($\tilde{\rho}_A$)=1, and det($\tilde{\rho}_A$)=0. Two things are immediately clear in this definition:

(1) *The* partial trace over qubit-B is equivalent to projecting the unit quaternion $\boldsymbol{t}$ onto the *k-axis* in the entanglement sphere: $bt \rightarrow bt_zk = bcos\chi k = x_4k$ (see Fig.7(b)). And

(2) $\tilde{\rho}_A$ indicates that the qubit-A quasi-Bloch sphere is the Riemann sphere with $x_1$-$b$-$x_0$ axes. This unit sphere intersects at its equator a complex plane which has $x_1$-real axis ($x$) and $b$-imaginary axis ($y$) with imaginary unit $\boldsymbol{t}$, and a point is given by $z=x+ty$ on this complex plane. Then, in terms of the stereographic projection, the point $(\theta_A, \phi_A)$ on the Bloch sphere is projected to a point $z$ on

the complex plane given by $z=\dfrac{x_1+bt}{1-x_0}=cot\dfrac{\theta_A}{2}e^{t\phi_A}$ (see Fig.7(a)). One the other hand, after tracing out qubit-B, $\rho_A$ indicates that the Bloch ball coordinate axes for the mixed-state single qubit-A are $x_1$-$x_4$-$x_0$ where the complex plane has $x_1$-real axis and $x_4$-imaginary axis with imaginary unit $k$. Then, the qubit-A Bloch coordinate is that of a Bloch ball given by $x_0^2+x_1^2+x_4^2=1-c^2$ where $c$ is the concurrence of the two-qubit state before taking the partial trace. In this single-qubit mixed-state Bloch ball, a constant concurrence corresponds to a sphere of radius $\sqrt{(1-c^2)}$. Using the angular parameters in this paper, the Bloch coordinate would be ($sin\theta_A cos\phi_A$, $sin\theta_A sin\phi_A cos\chi$, $cos\theta_A$). For a fixed angle $\chi$, this gives a flattened sphere, flattened along the imaginary-axis by factor $|cos\chi|$ relative to the unit 2-sphere. On the $x_0x_1$-plane, it has a unit radius (see Fig.7(c)). These relationships are illustrated in Figure 7. Finally, Figure 8 summarizes the maximally entangled states and the separable states on the qubit-A sphere and the entanglement sphere.

## VI. SUMMARY

A two-qubit pure-state Bloch sphere model was proposed and some examples of two-qubit states and gates were presented. The model consists of three unit 2-spheres (Bloch spheres of each qubit and a nonlocal, entanglement sphere) and a phase factor. This model can consistently represent both separable and entangled two-qubit pure states except one notable exception, $|\Psi_{AB}\rangle=|1\rangle_A\otimes|\psi_B\rangle$, in which case, the states can be represented directly, without following the procedure presented here. The model is presented in terms of qubit-A basis. This three sphere model seems to be able to represent all two qubit pure states and two qubit gates geometrically, enabling visualization and intuition. The main result is given by Eq.(24), which has a shortcut in Eq.(33). A detailed step-by-step procedure of parameter determination was given as the **step-1** through **step-4** in section-IV.

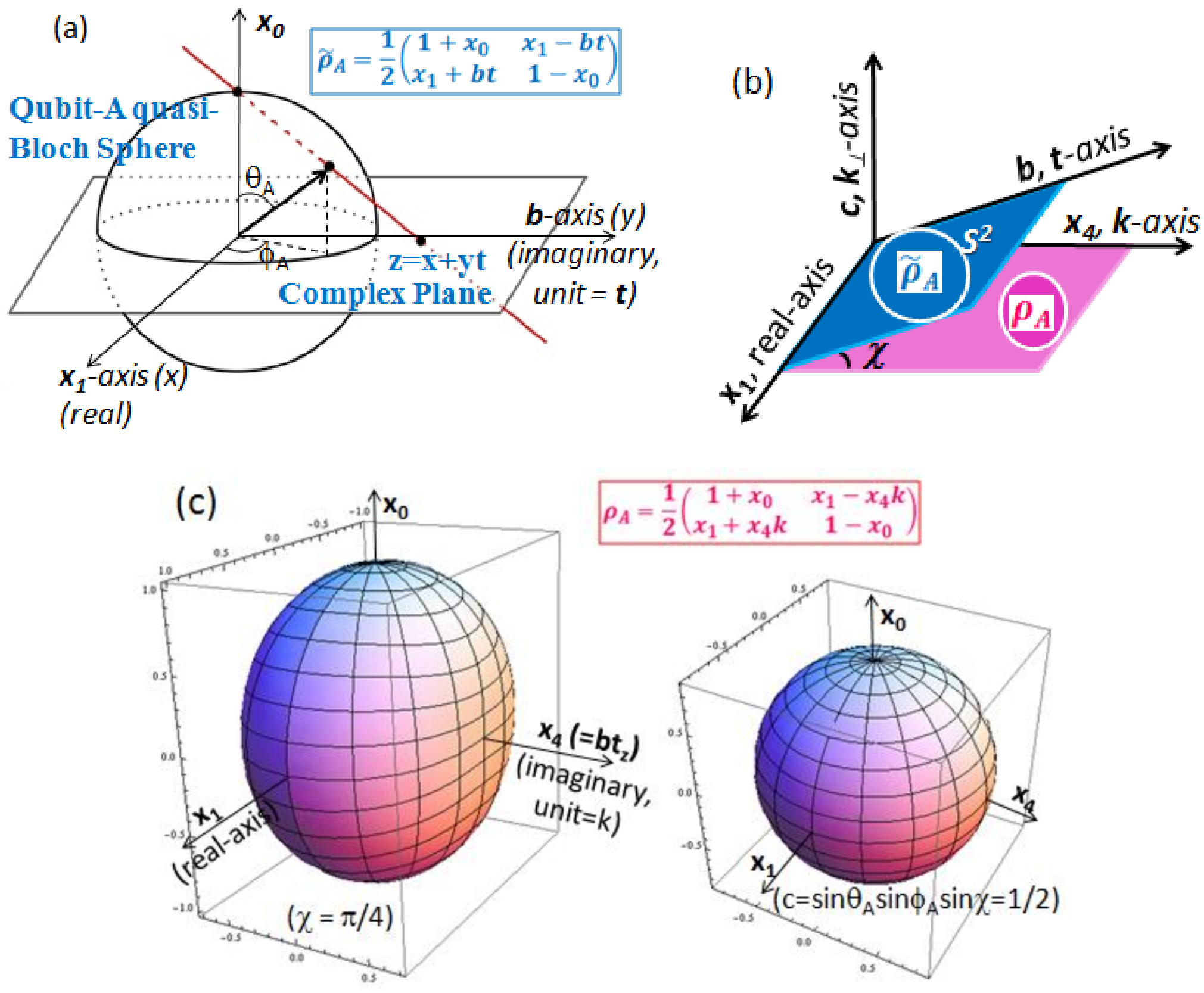

Figure 7 An illustration of Eq.(49). (a) The quasi-density matrix $\tilde{\rho}_A$ and quasi-Bloch sphere of qubit-A. (b) The complex planes. Partial trace over qubit-B projects the $\boldsymbol{t}$ imaginary-axis ($\boldsymbol{b}$-axis) to the $\boldsymbol{k}$ imaginary-axis ($\boldsymbol{x_4}$-axis), shrinking Bloch sphere(ball) in this direction by a factor $|cos\chi|$. The $\boldsymbol{k_\perp}$ imaginary-axis ($\boldsymbol{c}$-axis) is the concurrence axis (where $k_\perp=e^{k(\xi-\pi/2)}j$ , see Eq.(27)). (c) After tracing out qubit-B, the reduced density matrix $\rho_A$ is plotted for a constant $\chi$ (at $\chi=\pi/4$) and for a constant concurrence (at $c=1/2$).

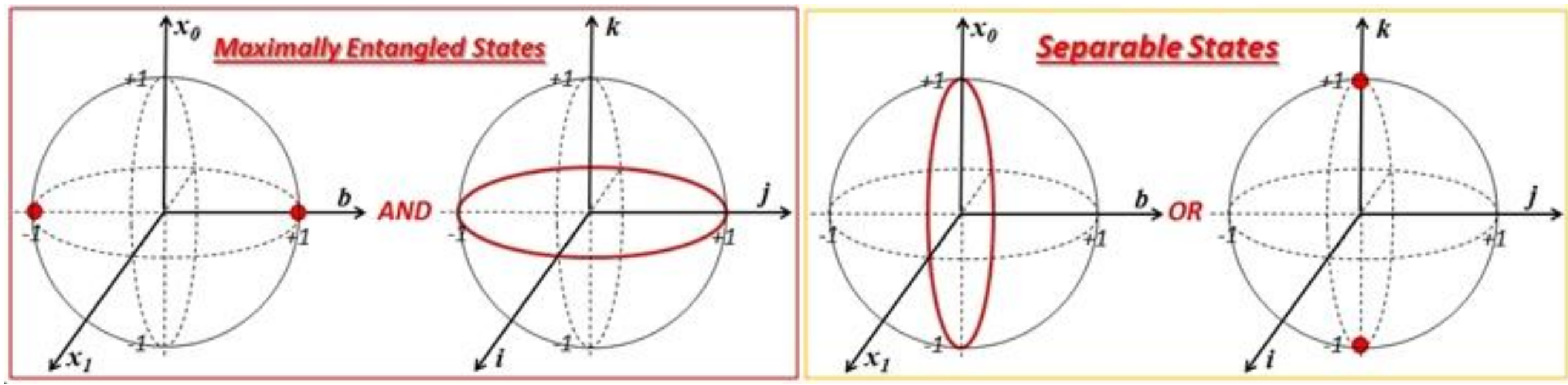

Figure 8 The maximally entangled states and the separable states are summarized on the qubit-A sphere and the entanglement sphere. This summary is consistent with the examples given in the Tables I, II and III.